\documentclass[useAMS,usenatbib,usegraphicx]{mn2e}
\usepackage{color}
\title[$\eta$ Carinae events]{The periodicity of  the  $\eta$ Carinae events\thanks{Based partially on data collected at the OPD-LNA/MCT}
\thanks{Based partially on data collected at ESO telescopes} 
\thanks{Based partially on data collected at Casleo Observatory}
\thanks{Based partially on data collected at Magellan Telescopes}
\thanks{Based partially on data collected at CTIO}
}
\author[A. Damineli, et al.]{A. Damineli$^1$\thanks{e-mail:damineli@astro.iag.usp.br}, 
M. F. Corcoran$^{3,4}$, D. J. Hillier$^{2}$, O. Stahl$^5$,  R. S. Levenhagen$^1$, 
\newauthor
 N.V. Leister$^1$, J. H. Groh$^1$, M. Teodoro$^1$,  J. F. Albacete Colombo $^6$, F. Gonzalez$^7$, J. 
\newauthor
 Arias$^8$, H. Levato$^7$, M. Grosso$^7$, N. Morrell$^{9}$,  R. Gamen$^{7}$, 
 G.Wallerstein$^{10}$,V. Niemela$^{11}$,\\ 
$^1$Instituto de Astronomia, Geof\'{\i}sica e Ci\^encias Atmosf\'ericas, 
Universidade de S\~ao Paulo, Rua do Mat\~ao 1226,\\ Cidade Universit\'aria,
 S\~ao Paulo, 05508-900, Brazil\\
$^2$Department of Physics and Astronomy, University of Pittsburgh,
 3941 O'Hara Street, Pittsburgh, PA 15260, USA \\
$^3$CRESST and X-ray Astrophysics Laboratory, NASA/GSFC, 
Greenbelt, MD 20771, USA\\
$^4$  Universities Space Research Association, 10211 Wincopin Circle, Suite 500 Columbia, MD 21044, USA\\
$^5$ZAH, Landessternwarte, K\"{o}nigstuhl 12, D-69117 Heidelberg, Germany\\
$^6$Facultad de Ciencias Astronomicas y Geofisicas de La Plata (FCAGLP)\\
$^7$Complejo Astronomico El Leoncito, Casilla de Correo 467, San Juan, Argentina\\
$^8$Departamento de Física, Universidad de La Serena, Chile\\
$^{9}$Las Campanas Observatory, Carnegie Observatories, Casilla 601, La Serena, Chile\\
$^{10}$Department of Astronomy, University of Washington, Seattle, WA 98195, USA\\
$^{11}$In memoriam\\
}

\begin{document}
\voffset=-3.0pc 

\date{Accepted XXXX XXX XX. Received YYYY YYY YY; in original form ZZZZ ZZZ ZZ}

\pagerange{\pageref{firstpage}--\pageref{lastpage}} \pubyear{2007}

\maketitle

\label{firstpage}

\begin{abstract}

Extensive spectral observations of $\eta$ Carinae over the last cycle,
and particularly around the 2003.5 low excitation event, have been obtained.
The variability of both narrow and broad lines, when combined with data taken from two earlier cycles,
reveal a common and well defined period.  We have combined the cycle lengths derived from 
the many lines in the optical spectrum with those from broad-band 
X-rays, optical and near-infrared observations, and obtained 
a period length of P$_{\rmn{pres}}~=~2022.7\pm1.3$ d.

 Spectroscopic data collected during the last 60 years yield an  
average period of P$_{\rmn{avg}}~=~2020\pm4$ d, consistent 
with the present day period. The period cannot have changed by more 
than $\Delta$P/P~=~0.0007 since 1948.  This confirms 
the previous claims of a true, stable periodicity, and gives strong  
support to the binary scenario. We have used the disappearance of
the narrow component of He\,{\sc i} 6678 to define the epoch of the Cycle 11 minimum, 
$T_0=$JD 2,452,819.8.  The next event is predicted to occur 
on 2009 January 11 ($\pm$2 days).
The dates for the start of the minimum
in other spectral features and broad-bands is very close to this date,
and have well determined time delays from the He\,{\sc i} epoch.  
\end{abstract}


\section{Introduction}\label{introduction}

 $\eta$ Carinae is one of the most luminous
stars in the Milky Way and contains many mysteries. 
It has been attracting attention since the 1820s, when it suffered large 
brightness fluctuations, culminating with the giant eruption that ejected the Homunculus in 1843. 
The star faded to naked eye invisibility, and after the discovery of the supernovae in the XIXth century
it was classified as a slow supernova.  However, around 1940, it started to brighten again, indicating 
that the star was only hidden by dust, not destroyed. 

The spectrum is rich in emission lines of low excitation species: H\,{\sc i}, Fe\,{\sc ii}, [Fe\,{\sc ii}], 
[Ni\,{\sc ii}], Ti\,{\sc ii}, etc. \citep{T53}; after 1944 \citep{b9} high excitation forbidden lines 
 of [Ne\,{\sc iii}], [Ar\,{\sc iii}], [S\,{\sc iii}], and [Fe\,{\sc iii}] can also be readily identified
  \citep[see also][and references therein]{b29}. Today we know that the narrow lines 
  (forbidden and permitted) are emitted in the Weigelt blobs \citep{weigelt86}, 
  at $\sim0.3$~arcsec from the central star \citep{b6} and the broad emission lines are formed 
  in the wind of the central object \citep{HA92_eta,b6}. The combination of high and low excitation 
 lines in the same object, however, was paradoxical.

\begin{figure*}
\vbox {\vfil
\begin{tabular}{cc}
\resizebox{17cm}{!}{\includegraphics{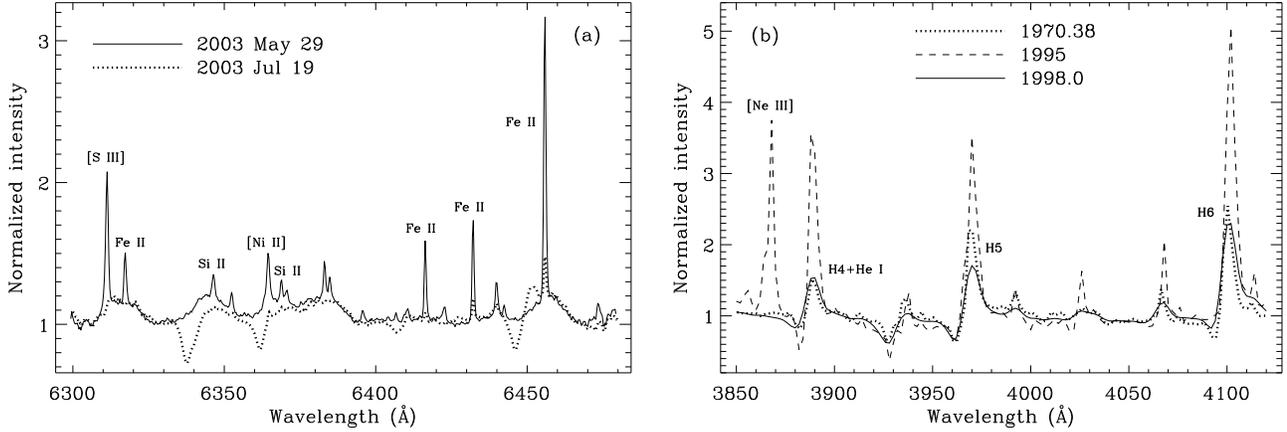}}
\end{tabular} 
\caption{\label{highlow} Sample of the $\eta$ Car spectra at high and 
low excitation states. (a) high resolution red spectrum showing the disappearance of 
narrow emission lines and 
strengthening of P Cygni absorption components during the minimum. (b) 
low resolution blue spectra showing the high state in 1995 and two spectra during the minimum taken 
in 1997 December 31 and 1970 May 17. 
  }\vfil}
\end{figure*}

A key to understanding this interesting 
object  was found recently through the study of the variability 
of the high excitation lines.
The high excitation forbidden  lines disappeared in 1948, and again in 1965, 1981, 
1987 and 1992.  These `spectroscopic events' \citep{b9,b17,b21,b25} 
or `low excitation events' \citep{b29} were believed to be part of 
S~Doradus cycles, commonly seen in other LBV stars similar to eta Car.
 This interpretation seemed to be supported by the 
 He\,{\sc i}~$\lambda$10830 line which went to minimum \citep{b3} 
 when the near-infrared light curve went to maximum \citep{b23}.
  The maxima in the near-infrared light curves
   were not truly periodic and the length of the quasi-period was 
   different for different pass-bands.
   However, the spectroscopic events were demonstrated to be 
  periodic \citep{b5}, in contrast to the incoherent character of the S Dor 
  oscillations.  \citet{b4} and others proposed 
   a binary model with a highly eccentric orbit, a hotter secondary 
   component and a strong wind-wind collision (WWC). 
   Binarity is interesting as it potentially allows the direct measurement 
of the mass of the stars, their most fundamental parameter. 
    The binary scenario has provided a 
    framework for understanding the star and provided guidelines 
    for fruitful observations, although some prefer a model
in which there are periodic shell ejections \citep{b15}.
In Fig. \ref{highlow} we present examples of high and low excitation state spectra of $\eta$~Carinae.

The observation of an event in 1997.95, as was predicted, brought 
more confidence to the true periodic nature of the variation \citep{b5}.
 \citet{b8} used archival spectra to identify three previously unreported events, in 1953, 1959 and 1970, which also fit the 5.5-yr period.
  Moreover, those authors discovered that 
 the dips on top of the broad quasi-periodic near-infrared maxima were truly periodic
  and correlated  with the behavior of the high excitation lines. An extensive
  X-ray monitoring campaign was started in 1996 with the {\em RXTE} 
  satellite and revealed deep minima in 1997.95 and 2003.49 which coincided 
  with the minima seen at other wavelengths \citep{b2}. X-ray observations 
  inside and outside the minimum performed with {\em Chandra} and {\em XMM}
   furnished details on the column density ($N_{\rm{H}}$), temperature 
   and chemical composition of the colliding wind shock \citep{b11}. 
   \citet{b22} showed that the optical light curve displays periodic dips 
   like those in the near-infrared and \citet{b14} reported a very detailed
    light curve in the $B$, $V$, $R$ and $I$ bands for the 2003.49 event. The events were
     recorded also at radio-cm \citep{b7} and radio-mm \citep{b1}, but no 
     specific value to the period length was reported for those wavelengths. 

 Many other features vary periodically in intensity and radial velocity, like the
broad emission and P Cygni absorption components,  and can also be used
to derive the period length. 
One of them is He\,{\sc ii}~$\lambda$4686 discovered by \citet{b20}, which  
raises and drops just before minimum faster than any other  eature over the entire spectrum. 
Although faint (EW$<2$~\AA) it was frequently monitored with high signal/noise along the last event.
Unfortunately, it was observed only occasionally in the previous events, precluding its use to 
measure the period. This spectral line deserves better monitoring in future events, not only to 
improve the accuracy of the derived period, but also because it is the highest excitation feature observed at optical wavelengths, and its origin remains a mystery.


 To facilitate discussion we 
   label the events by numbers as described by \citet{b10}: number one (\#1) is 
   assigned to the event observed in 1948 by Gaviola, so that the event of 
   2003.49 is \#11. We define {\em cycle} as the time interval between 
   the starting of two consecutive minima, so that cycle~\#9 started 
   at the 1992.42 minimum and finished when cycle~\#10 was starting in
    1997.95. Because of observational reasons, that will become clear 
    later in this paper, the starting point of a cycle is
     defined by the disappearance of the 
He\,{\sc i}~$\lambda6678$ narrow line component. With this definition, phases 
along the cycle are defined in a unique way for all measured quantities.

  The paper is organized as follows.  We present in section \ref{observations} the observations;
   in \S \ref{zero} the definition of the phase 0 of the minimum;
   in \S \ref{period} the determination of the period length;
   in \S \ref{stability} the stability of the period; 
   in \S \ref{eruption} the relation between the sharp peaks during the giant eruption 
                  and periastron passages; 
   and in \S \ref{discus} the discussion and conclusions. 

\section{Observations and measurements}\label{observations}

The majority of the ground-based observations presented in this paper came 
from a monitoring campaign started in 1989 at the Coud\'{e} focus of the 
1.6-m telescope of Pico dos Dias Observatory (OPD-LNA/Brazil). The observational 
setup at OPD was kept essentially unchanged through the campaign: a dispersion
 grating with 600 l/mm, entrance slit width $\sim1.3$ arcsec, exposure time 
 $\sim 5$ s in H$\alpha$ increasing to $\sim15$ min at 3500 and 10800~\AA. 
 Spectra were extracted along $\sim2$ arcsec in the spatial direction and no
  measurable differences in line intensities were seen when changing the 
  extraction size by a factor of 2. Three different CCDs have been used, 
  with resolving powers R = 25 km~s$^{-1}$ (0.25 \AA~pixel$^{-1}$) at 
  H$\alpha$ in 2003 and R = 50 km~s$^{-1}$ (0.39 \AA~pixel$^{-1}$) in the 
  preceding years. On some occasions, a $1024\times1024$ Hawaii detector 
  was used to observe the He\,{\sc i}~$\lambda$10830 line, delivering a 
  spectral resolution R = 40 km~s$^{-1}$ (0.65~\AA~pixel$^{-1}$). On
  other occasions, spectra
   of this line were taken at R = 15 km~s$^{-1}$ with a thinned CCD. After
    correcting for fringes and degrading the spectral resolution, these 
    spectra were almost identical to those collected with the infrared array 
    at the same date.
     Observations in 1992 and 1997/8 were done with a thick CCD that was 
     almost free of fringes, but had a low sensitivity in the blue, which 
     explains the poor coverage of important lines in that spectral range. 
     For wavelengths longer than 6500~\AA, telluric absorptions and fringes 
     (in thinned CCDs) were removed by using templates constructed from spectra 
     of bright early type stars ($\theta$ Carinae, $\zeta$ Ophiuci or $\zeta$ Puppis) 
     observed immediately after or before $\eta$ Car.

\begin{figure*}
\vbox {\vfil
\resizebox{17cm}{!}{\includegraphics{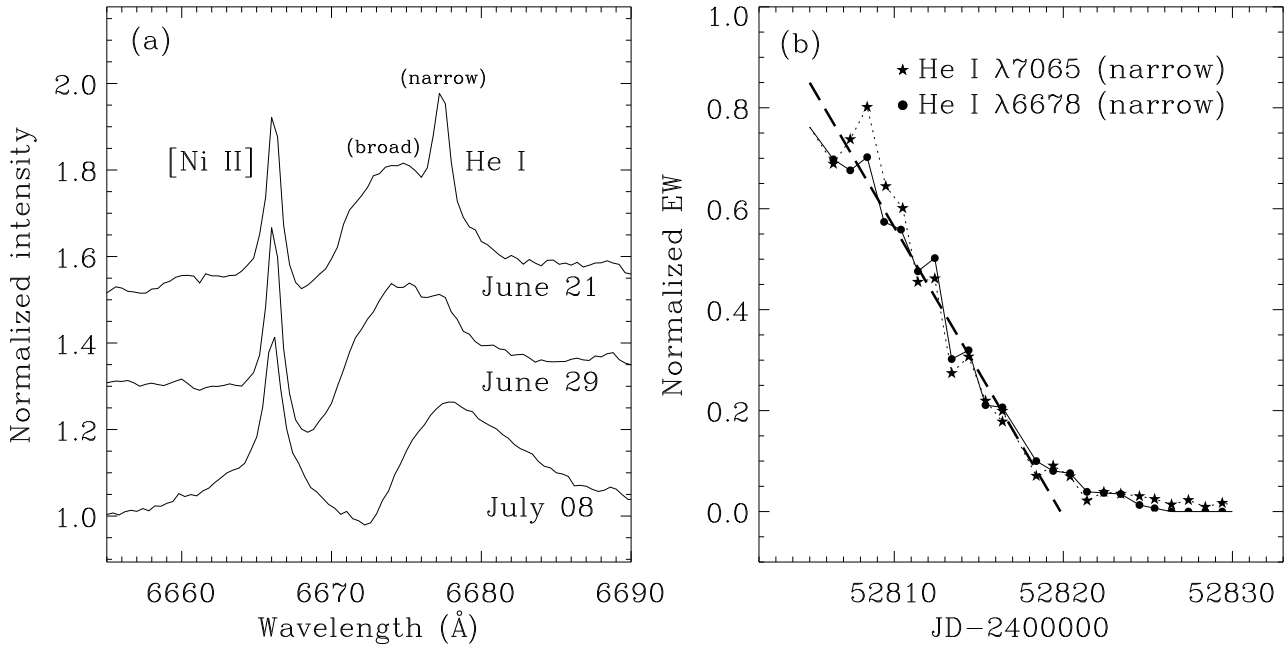}}
\caption{\label{phasezero} Definition of phase 0 -- (a) He\,{\sc i} $\lambda6678$ 
line profiles observed along 17 days in 2003, showing the disappearance of the narrow 
component. (b) The fading phase, showing our method to derive the phase 0  (JD~2,452,819.8) of 
He\,{\sc i} $\lambda6678$ and $\lambda7065$ narrow line components. } \vfil}
\end{figure*}

For the 2003 event, we also used spectra taken with the spectrographs REOSC (R = 25 km~s$^{-1}$) 
and EBASIM (R = 7 km~s$^{-1}$) attached to the 2.15-m CASLEO telescope (Argentina), and spectra 
taken at CTIO with the 4.0-m Echelle Spetrograph (R = 8 km~s$^{-1}$) and at Magellan 
with MIKE Spectrograph (R = 12 km~s$^{-1}$). For the 1997/8 event, we used 
spectra collected at La Silla/ESO with CAT-CES (R = 12 km~s$^{-1}$). For the 1992 event,
 we also used spectra collected with the FLASH/HEROS spectrograph attached to the 50-cm 
 telescope (ESO/Chile) with a fiber diameter $\sim5$~arcsec and spectral resolution 
R = 12 km~s$^{-1}$. On several occasions we used FEROS spectrograph attached to the 1.52-m telescope 
at La Silla to cover the entire optical window at resolution R = 12 km~s$^{-1}$.

Before measuring the spectral features,
we degraded the spectra to a dispersion of 0.39~\AA~pixel$^{-1}$. This step was not really 
necessary but it helped facilitate the adoption of the same limits between the narrow
and broad 
components, and positioning of the stellar continuum, when measuring the spectra.
Since we adopted the observations collected at LNA Observatory as a reference, 
we added data from other sources only in the case where they merged smoothly to the line intensity curve.
 This criterion was fulfilled by almost all ground-based observations, 
 confirming our expectation that slit widths in the range $1-3$ arcsec width 
would give the same results independent of the position angle (P.A.) of the slit. This happens because the 
main emitting region is smaller than $1$ arcsec and has a huge contrast to the
surrounding Homunculus nebula and also because the seeing fwhm is 
larger than $1$ arcsec, smearing out the emitting region. 
In a forthcoming paper (on the long term behavior of the spectral lines) we will 
present the complete list of observations from the entire campaign and a table with 
individual measurements. Fig.~\ref{highlow}a displays spectra 
representative of the high
 and low excitation states, showing the disappearance of the high excitation lines 
 and enhancement of P Cygni absorption  profiles during the minimum.  Fig.~\ref{highlow}b 
 shows spectra in the blue for the high excitation state of 1995 and for the 
 low excitation state of 1997 and 1970 (see also \citet{b29} for the full spectral range 3850--11000 \AA). 

Spectra from the Space Telescope Imaging Spectrometer (STIS) on the 
{\em Hubble Space Telescope} are available for the 2003.49 and 1997.95 events, 
though for consistency we do not include them here since the slit width is 
much narrower than used in the ground-based observations, sampling only a 
part of the inner circumstellar nebulosity. These data are of course important 
for disentangling stellar from circumstellar variations, and have been more 
fully described in \citet{nielsenetal07}, \citet{krister07}, \citet{gull06}, 
and \citet{kd05}. Since the wind of the primary star is resolved by the 
STIS slit and the slit's position angle varied in different visits, care 
must be taken when comparing  line profiles from different epochs. 
This applies to the lower excitation transitions, formed 
far from the central source(s) that may be subject  to spatial asymmetries.

All the data processing and measurements were done in the standard way using 
IRAF packages.  Narrow lines were modeled by Gaussian fitting and deblended from 
the broad components.
Since they are seated on top of broad line profiles, which are themselves variable, 
  we referred their equivalent widths (EW) to the local stellar continuum, in order that 
these measurements correspond to line flux normalized to the local stellar 
continuum, instead of classical  equivalent width. As in the case of EW, this kind of
measurement is translated into line flux when multiplied by the stellar continuum flux.
Because of this, we use the simple designation of equivalent width in place of normalized 
line flux.  
 Broad line emission profiles were separated from the narrow components, when they existed, 
 and their  equivalent widths and baricenters (for radial velocities) were measured by 
direct integration along the line profile. 
  Radial velocities are in the heliocentric reference system.

   It is difficult to attribute errors to single measurements, as the main source is 
    systematic, not statistical. The spectra were well exposed, in order that photon noise
   is very low, except in the violet region. The major source of error is linked to the stellar continuum, because of 
   line blendings and  changes in relative intensity of line/continuum, as the seeing 
   changes and smears out the central source of emission lines.   
  The random errors can be judged by the smoothness of the 
   curves in line intensity and the plots show that they are small, in general comparable to 
   the size  of the symbols in the figures. We minimized the errors  by over-plotting the spectra 
     and pointing the cursor always in the same postition. We must warn, however, that this
     procedure  does not eliminate  the systematic errors.

\section{Defining phase 0 for the spectroscopic event}\label{zero}

A simple method to measure the periodicity of the events is through the 
disappearance of spectroscopic features like the high ionization lines or
 the narrow components of He\,{\sc i} (Fig.~\ref{phasezero}a). In practice
  this is difficult because of the following: a)~the time 
  sampling has been too coarse to pick up the exact time when the feature 
  disappears; b)~spectroscopic features reach minimum at different times; 
  c)~minima are usually reached asymptotically for many important spectral 
  features, often taking up to a week to disappear completely; d)~when the 
  line EWs are less than $\sim100$~m\AA, they are difficult to measure, 
  unless the spectra have very high signal-to-noise ratio (S/N). 
  In addition to producing a large uncertainty in the epoch of the minimum,
   faint features may not be directly connected to the emitting region, 
   but can be light echoes that fall inside the slit aperture. Moreover, 
   in some cases a very faint blended line remains in emission through the 
   minimum, as in the case of [Ar\,{\sc iii}]~$\lambda7135$. 

In order to minimize these problems we 
restricted our analysis to the phase of steep decline, which lasts for 
$\sim2$ weeks, starting $\sim3$ weeks before complete disappearance. 
We performed a linear fit, and extrapolated it to zero intensity to 
determine the time of minimum (Fig. \ref{phasezero}b). This 
procedure is much more robust than other techniques, since it does not 
require a dense time sampling along the minimum.  It is relatively insensitive 
to the S/N of the spectrum, and is easily reproducible by other
observers. The epoch of minimum, i.e. phase 0 (the 
starting point of the deepest part of the minimum) for the He\,{\sc i} 
$\lambda6678$ narrow line component derived by this method is T$_0$~=~JD~2,452,819.8 
(2003 June 29 or 2003.491). Since He\,{\sc i} $\lambda6678$ is strategic for the 
spectroscopic event (it has a long observational history, shows clear, easily
 measured variability and lies in a spectral range with good CCD efficiency)
 we chose it for our definition of phase 0. 

There are two situations for which it is useful to find signatures that 
indicate the time of phase 0: when examining non-calibrated historical
spectra or when trying to track the evolution of an event during a monitoring 
campaign. As the high excitation lines are much more variable than 
the lower excitation lines, and because the spectrum has plenty of lines,
it is relatively easy to find line pairs that interchange peak 
intensity ratio with time. A high excitation line, as the minimum approaches, decreases 
until its peak is equal in strength to that of some nearby low 
excitation line (in general Fe\,{\sc ii} or [Fe\,{\sc ii}]), and
we record the date when this occurs. The faster the high
excitation line varies the more accurate is the determination of the
time of change in the line ratio. This happens for dates close to phase
zero, when the variability is high, but we were able to find good line
pairs up to three months before phase 0 and almost two years after.

\begin{figure*}
\vbox {\vfil
\begin{tabular}{cc}
\resizebox{18cm}{!}{\includegraphics{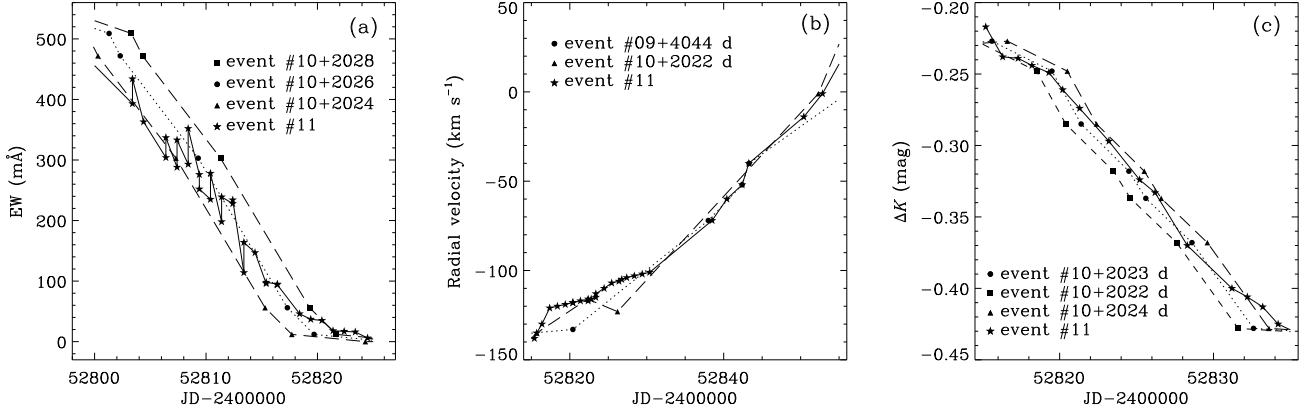}}
\end{tabular}
\caption{\label{periodfig} Period length: (a) from the EW He\,{\sc i}~$\lambda6678$ 
narrow line component along the latest 2 events; (b) from radial velocity of the He\,{\sc i}~$\lambda6678$  
broad component in the last 3 events; (c) from the $K$-band 
by using the method of folding the 2 latest events and minimizing the residuals.
}
\vfil}
\end{figure*}

We display in Table \ref{pkratio} the time in days for the inversion in peak intensity 
ratios, relative to phase 0. Negative values represent dates before phase 0 and 
positive values dates later than phase 0. Entries in column 2 are for the fading 
phase and in column 3 for the recovering phase, except for He\,{\sc i}~$\lambda$10830,
which displays the two ratio inversions in the fading phase. Times are shorter in 
column 2 than in column 3 due to the fact that the fading phase is fast and the 
recovering phase is slow. From an examination of data for the last three cycles,
we found that times in column 2 are accurate to $\sim15$ per cent and in 
column 3 to $\sim25$ per cent.

\begin{table}
\centering
\caption{\label{pkratio}Time delays in days, relative to phase 0, when 
the intensity of line peaks change ratio.}
\begin{tabular}{lll}
\hline
Line ratio    &   change to $<1$ & change to $>1$ \\ 
\hline
He\,{\sc i}~10830 R/V* peaks & -105 & -4  \\
He\,{\sc i}~4471/[Fe\,{\sc ii}]~4475 & -46& +550 \\
 
[S\,{\sc iii}]~6312/Fe\,{\sc ii}~6317 & -15& +442 \\
 
[Fe\,{\sc iii}]~4658/[Fe\,{\sc ii}]~4640 & -17& +358 \\
 
He\,{\sc i}~7065/[Fe\,{\sc ii}]~7171& -9& +148 \\

He\,{\sc i}~6678/[Ni\,{\sc ii}]~6666  & -9& +145 \\
 
[Fe\,{\sc iii}]~4658/[Fe\,{\sc ii}]~4475 & -8&+250  \\
 
[Fe\,{\sc iii}]~4658/[Fe\,{\sc ii}]~4665 & -5& +168 \\
 
He\,{\sc i}~5876/Na\,{\sc i}~5890 & 0& +18 \\
 
[N\,{\sc ii}]~5754/[Fe\,{\sc ii}]~5746  & +1& +79 \\
 
Fe\,{\sc ii}~8490/Fe\,{\sc ii}~8499 & +7& +40 \\
\hline
\multicolumn{3}{l}{* `red'  (R) and `violet' (V) peaks intensity ratio}
\end{tabular}
\end{table}

In Table \ref{pkratio} we have also listed the line He\,{\sc i}~$\lambda$10830. It has a 
double peak, like in classical Be stars. The V (`violet') and R (`red') peaks 
are variable, both in intensity and in their relative strength. For almost the 
entire 5.5-yr cycle,  $\rmn{R}>\rmn{V}$. As the minimum approaches, 
the R peak starts decreasing faster than the other, in such a manner that 
105 days before phase 0 they reach V = R, changing to $\rmn{R}<\rmn{V}$ subsequently. 
The $\rmn{R}<\rmn{V}$ state lasts for almost three months when the rate of fading 
of the R peak slows down and the V peak starts falling fast. Just four days 
before phase 0, the peaks again reach V=R and return to $\rmn{R}>\rmn{V}$.

\section{The period length}\label{period}

There are a number of ways to measure the period length; the best one for 
spectroscopic data is based on the He\,{\sc i} narrow line components. The equivalent width 
of this feature remained relatively constant at $\sim$1500~m\AA~for most of the cycle \#10.
 About three weeks before phase 0, it began to change fast, declining 
 by $\sim25$~m\AA\ day$^{-1}$. We used this segment of the line intensity 
 curve to measure the period, applying a scheme of epoch folding and minimization 
 of differences similar to that used by \citet{b2}. Since we sampled better the 
 fading phase to the minimum, it was sufficient to shift this piece of the line 
 intensity curve from the event \#10 until it matched that of event \#11 (Fig.
  \ref{periodfig}a) to derive the period. We repeated the same procedure with the event \#9, getting the
   best fit for P~=~2026 days with an uncertainty of 2 days. A careful examination 
   of Fig. \ref{periodfig}a, however, indicates that the slope of the fading phase 
   was steeper during event \#10 than in event \#11. This is due to secular changing in the line intensity
   and this is the main source of errors in the period determination by this method.

\begin{table}
 \centering
  \caption{\label{length} Period length in days from different spectral regions}
  \begin{tabular}{ll}
  \hline
   Period $\pm$ error & spectral feature or pass-band \\
 \hline
  2026   $\pm$ 2   & He\,{\sc i}~6678 narrow component Eq. W.\\
  2024   $\pm$ 2   & $X$-rays \\
  2023   $\pm$ 1   & $J$-band \\
  2023   $\pm$ 1   & $H$-band \\
  2023   $\pm$ 1   & $K$-band \\
  2023   $\pm$ 2   & $L$-band \\
  2022   $\pm$ 2   & Fe\,{\sc ii}~6455 P Cyg abs. radial velocity\\
  2022   $\pm$ 1   & Si\,{\sc ii}~6347 Equivalent Width\\
  2022   $\pm$ 1   & He\,{\sc i}~6678 broad radial velocity\\
  2022   $\pm$ 1   & He\,{\sc i}~10830 Equivalent Width\\
  2021.5 $\pm$ 2   & $V$-band \\
  2021   $\pm$ 2   & Fe\,{\sc ii}~6455 P Cyg abs.  Eq. W. \\
  
  2022.7 $\pm$ 1.3 & average $\pm$ standard deviation   \\
\hline
\end{tabular}
\end{table}

The broad component of He\,{\sc i}~$\lambda6678$ also changes quickly before 
the minimum. Its radial velocity decreases slowly in-between the events, but three
 weeks before phase 0 it reverses the trend, and starts to increase. About 
 4 weeks after phase 0, the radial velocity increases at a rate of
  $\sim~5$km~s$^{-1}$ per day. The steep rise in the radial velocity 
  curve is useful to determine the period length, in the same way as 
  we have done for intensity of the narrow component. This 
  procedure might be more robust than using equivalent widths, 
  since radial velocities are much less affected by the secular variations 
  in intensity. By combining RVs from the last three events (\#9, \#10 and 
  \#11) we derive P~=~$2022\pm1$~day (Fig. \ref{periodfig}b). 

Other spectral features were observed, whenever possible, and some of them 
also were useful for measuring the period length. The total equivalent width 
of the He\,{\sc i}~$\lambda$10830 line recorded in the last three events 
gives P~=~$2022\pm1$~days. The equivalent width of the P Cygni absorption component of 
the Fe\,{\sc ii}~$\lambda$~6455\,\AA\ line (also using the last three events) 
gives P~=~$2021\pm2$ days and the radial velocity curve of the same line 
(events {\#}9 and {\#}11) results in P =~$2022\pm2$ days. The equivalent 
width of Si\,{\sc ii}~$\lambda$~6347\,\AA\ P~Cygni component (events {\#}9 
and {\#}11) results in P~=~$2022\pm1$~days.

These periods are in good agreement with P~=~$2024\pm2$~d derived from X-rays 
\citep{b2}, with P~=~$2023\pm3$~d from the $K$-band light curve \citep{b24}, 
and with P~=~$2021.5\pm0.5$~d from  $V$-band \citep{b22}. We re-derived the 
periods obtained by those authors, using their published data and applying the 
technique of folding the pre-minimum branch of two events, \#10 and \#11 in the
 case of \citet{b24} and {\#}7 and {\#}11 in the case of \citet{b22}. 
  In the case of $V$-band photometry, we
   combined the light curve of \citet{b22} with that of \citet{b14} in 
   order to get a better definition of the descending branch of the 2003.5 
minimum. The derived period was the same as published by \citet{b22}, but
with an uncertainty of 2\,d instead of 0.5\,d. In the case of near-infrared,
     \citet{b24} used the lower point in the $K$-band minimum. Our procedure 
     of minimization of residuals applied to the $JHKL-$band photometry (Fig.
      \ref{periodfig}c) gave the same period, but with a tighter constraint.
 It is encouraging to see that the period is robustly defined, 
      independent of the particular choices made by different authors during the
       measurements (Table \ref{length}).

Since there is no reason to suppose that the period length would depend on 
the particular technique used, we combined all these individual periods to get 
a mean value to the period. We call it the present day period (P$_\rmn{pres})$ 
to differentiate from that determined from historical observations. 
Since the systematic errors may be
more important than statistical errors, we report the uncertainty in the period
as a simple standard deviation.

\[
\rmn{P}_{\rmn{pres}} = 2022.7 \pm  1.3~\rmn{d} \hspace{0.2cm}  \rmn{or} \hspace{0.2cm} 
\rmn{P}_{\rmn{pres}} = 5.538 \pm 0.004~\rmn{yr}.
\]

Regarding the times of phase 0, there is no reason to expect that 
different features give the same epoch, since they are produced in a variety 
of regions -- in the stellar winds, in the WWC  and in the circumstellar material. Since we are 
dealing mostly with spectroscopic lines we define, for reasons
discussed earlier, phase 0 of the spectroscopic events from the disappearance 
of the He\,{\sc i} narrow line intensity. 
This yields the following ephemeris:

\[
\rmn{JD}(\rmn{phase~0}) = 2,452,819.8  \pm 0.5 + (2022.7 \pm  1.3 \rmn{d}) \rmn{E}.
\]

The uncertainty is only 0.07 per cent of the period length, which enables 
us to accurately predict the time of phase 0 of the next spectroscopic 
event: JD~2,454,842.5 $\pm$ 2 (2009 January 9--13).

\section{Stability of the period}\label{stability}

An important question is the long term stability of the period, since the companion 
stars are losing mass at high rates and tidally interact during the periastron
 passages. In addition, the primary star could be a fast rotator  -- as 
 indicated by its dense polar wind \citep{boekel05, weigelt07, SGH03_IR} -- and the 
 specific angular momentum may be 
 changing continuously. The coupling of rotational and orbital angular momentum 
 may lead to an increase in the orbital period. A period derived from data 
 encompassing many cycles may be hiding such variations. The only way to tackle 
 this question is by comparing the present day period (measured from the last 
 few events) with the average period (P$_{\rmn{avg}}$), derived from the events 
 recorded in the past 60 years. 

\begin{figure*}
\vbox {\vfil
\resizebox{17cm}{!}{\includegraphics{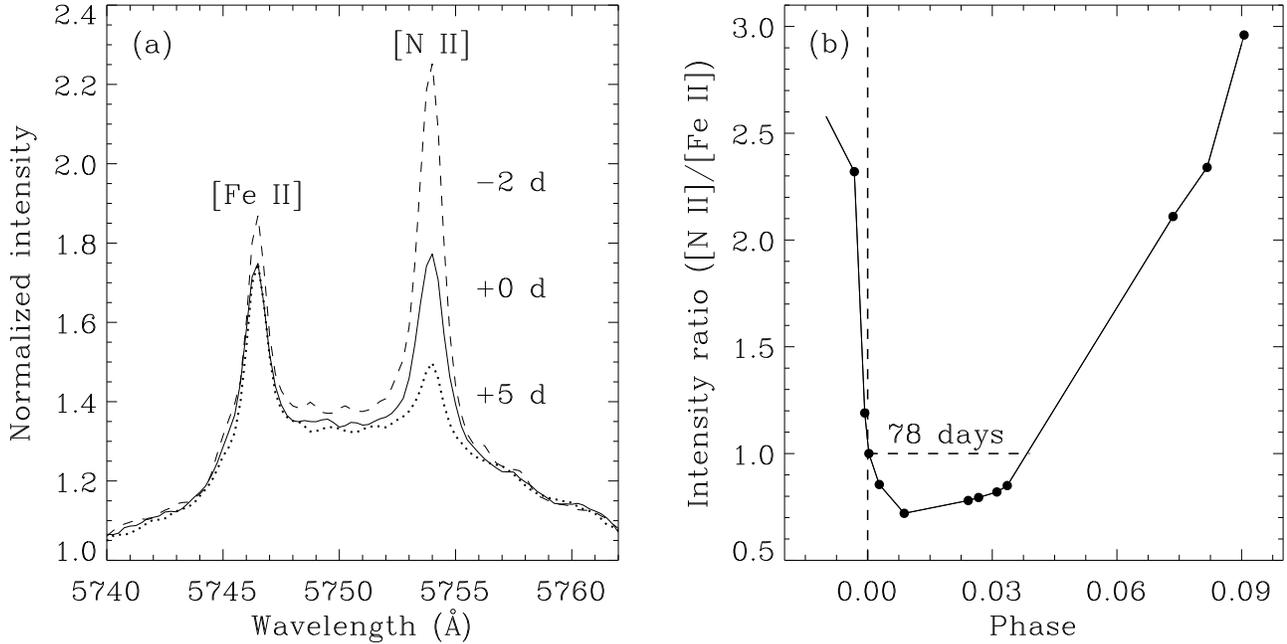}}
\caption{\label{n2f2} Relative intensities of [N\,{\sc ii}]~$\lambda5754$ 
and [Fe\,{\sc ii}]~$\lambda5746$ -- (a) Line profile variations. Labels 
at right of [N\,{\sc ii}] peak indicate days from phase 0 of the 2003.49 
event. (b) Ratio between line peaks, indicating that [N\,{\sc ii}]/[Fe\,{\sc ii}]$<1.0$ 
starts at phase 0 and lasts for 78 days during the minimum.}\vfil}
\end{figure*}

\begin{table}
 \centering
\caption{\label{paverage}Period derived from epochs of predicted and observed minima.}
\begin{tabular}{lllll}
\hline
Cycle & Pred. & Predicted&Observed &  P \\
 & JD (**) & Date&Date&  days\\
\hline
\#1 & 32,592.8& 1948 Feb 11 &1948 Apr 19 &$>$2015.9\\
\#2 & 34,615.5& 1953 Oct 26 &1953 Jun 28* &$<$2029.3\\
\#2 & "              &"                    & 1953 Dec 30 & $>$ 2009\\
\#3 & 36,638.2& 1959 Mar 11 &1959 May 14 &$>$2014.4\\
\#4 & 38,660.9& 1964 Sep 23 &1965 Feb 15 &$>$2001.9 \\
\#5 & 40,683.6& 1970 Apr 07 &1970 May 17 &$>$2016.0\\
\#7 & 44,729.0& 1981 May 04 &1981 May 21 &$>$2018.6\\
\#8 & 46,751.7& 1986 Nov 17  &1987 Jan 15 & $>$2003.1\\
\#9 & 48,774.40& 1992 June 01 &1992 May 31&$-$\\
\#10 & 50,797.10& 1997 Dec 15 & 1997 Dec 12 & $-$\\
\hline
\multicolumn{5}{l}{* Intermediate excitation, ** JD+2,400,000}
\end{tabular}
\end{table}

We looked at the events reported by \citet{b8}. In case the spectrum was in
 low excitation state, we assumed that it had just reached the minimum and so 
 was at phase 0. As a matter of fact, phase 0 must have occurred already 
 at some time before the observation, in order that a period based on that record
  would underestimate the period. The average period, or more precisely a lower 
  limit to it, is obtained from the time interval between that date and the last 
  recorded minimum (2003.49) divided by the number of cycles.  The derived 
  average periods are presented in column 5 of Table \ref{paverage}. We 
  excluded the last three events, since they were used to derive the present 
  day period. The dates of observed minima in Table \ref{paverage} (column 4) 
  were taken from \citet{b8}, except for the observation on 1970 May 17. This 
  spectrum (taken at CTIO) was recorded by Virpi Niemela and indicates that the 
  phase 0 occurred at least 8 days before Thackeray's observation reported by  
\citet{b8} . The CTIO 
  logbook reports that spectra were taken by Barry Lasker a day before phase 0 
  of the 1970 event (April 6), which could lead to a very tight constraint on the 
  period, but unfortunately we were not able to locate that spectral plate.

The observation made on 1948 April 19 gives P$_\rmn{avg}>2015.9$~d and that of 1981 
May 21 gives P$_\rmn{avg}>2018.6$~d. We can constrain the period length also from 
the other side. A maximum period may be derived when a particular 
observation was made before phase 0. This is the case for the observation 
made on 1953 June 28, when the star was approaching the minimum, but was still 
in an intermediate phase, which gives P$_\rmn{avg}<2029.3$~d. The average period 
is thus constrained to:

\[
2029.3 > \rmn{P}_{\rmn{avg}} > 2018.6~\rmn{d}.
\]

The stability of the period can be obtained from the difference between the 
present day period and the average period, taking into account that 
P$_\rmn{avg}$ refers to half of the cycles involved. The spectrum taken in 
1953 June 28 indicates that the period cannot have decreased by more than
 1.4~d~cycle$^{-1}$ and that of 1948 April 19 implies that it cannot have 
 increased by more than 1.4~d~cycle$^{-1}$. 

We have another way to constrain the average period using quantitative 
information of the first event in 1948. \citet{b9} reported that 
[N\,{\sc ii}]~$\lambda5754$ was fainter than [Fe\,{\sc ii}]~$\lambda5746$ , 
which places the date of the observation in a particular range inside the 
low excitation event. Examination of recent events indicates that before 
phase 0, [N\,{\sc ii}] is much stronger than the neighboring [Fe\,{\sc ii}] 
line. The [N\,{\sc ii}] line decreases quickly, in contrast to [Fe\,{\sc ii}]
 which undergoes small and slow changes. Both features reach equal intensity 
 0.7 days after phase 0, as can be seen in Fig,~\ref{n2f2}a, where 
 variations during event \#11 are displayed. The [N\,{\sc ii}] line remains 
 fainter than [Fe\,{\sc ii}] for a subsequent 78 days. This can be seen in Fig. 
 \ref{n2f2}b, which combines measurements made in the last three events. 
 The ratio of these two lines is not sensitive to the slit width or to the 
 spectral resolution, as long as they are kept $<4$ arcsec, or $\rmn{R}>2000$,
  respectively. The fact that the 1948 observation was done $<78$ days later than 
  phase 0, combined with the epoch of the 2003.49 minimum results in:

\[
\rmn{P}_{\rmn{avg}} = 2020 \pm  4~\rmn{d}.
\]

This  period is compatible with that derived in the present day data. Taking into
 account that the average was taken between 10~cycles, it could not have 
 changed by more than 1.5~d~cycle$^{-1}$, in close agreement with the 
 1.4~d~cycle$^{-1}$ derived before. The constraint to the period change is:

\[
-0.0007 < \Delta \rmn{P/P} < +0.0007.
\]

This confirms previous claims of strict periodicity by \citet{b3, b5} 
and \citet{b8} implying that the low excitation events are only understandable in the
binary scenario. It is also consistent with the period change expected
because of mass-loss from the primary star. Simple considerations show that 
$\dot\rmn{P}/\rmn{P} = \alpha \dot M/M$
where $\alpha$ is a constant of order unity  \citep{Kha74}. Ignoring changes
in eccentricity we find:

\[
\dot \rmn{P}= 0.11 \alpha \left( {\dot M \over 10^{-3}\,M_\odot\,\rmn{yr}^{-1}} \right)
\left({100 M_\odot  \over M}\right) \rmn{days/cycle}.
\]

\noindent
This is fully consistent with the observed upper limit.

\section{Were the peaks in the giant eruption produced by
 periastron passages?}\label{eruption}

\citet{b3} pointed out that the three most pronounced peaks observed 
in 1827--1843 were in close coincidence with the predicted times of 
phase 0, when using the period P~=~2014~d. \citet{b26}  also
noted that the `Lesser Eruption' which began on 1887.5 was within a 
few months of phase 0. When using the new period, derived in this work (P~=~2022.7~d), 
the 1827.087 peak is now at phase 0.15; that of 1837.967 is at
 phase 0.13: and that of 1843.3 at phase 0.05 \citep{b26}, while the 
 1887.5 event corresponds to a phase of 0.96. The correlation between 
 peaks and the start of the spectroscopic events worsens with the new 
 ephemeris. However, an exact correlation between the times of the peaks 
 and phase 0 is not really relevant, since the time sampling of the
  visual light curve was not very dense and the real maxima could have
   been missed by the observers. Moreover, if the mechanism that 
   produced those peaks is the same as the one that produces the broad
    maxima observed presently in the near-infrared light curve, the 
    lack of coincidence with phase 0 would not be a surprise. As 
    reported by \citet{b24}, the $JHK$- light curves present maxima 
    around phase 0, but not in exact coincidence. These near-infrared peaks are 
    quasi-periodic and may remain in high state for up to 3 years,
     depending on the wavelength. Although the near-infrared light-curve is
anti-correlated  with the radio flux, it still can be explained as 
     free-free emission if optical depth effects are taken into account \citep{b24}. 

In the $V$-band, the maxima associated with the periastron are inconspicuous, 
as compared to those in the near-infrared. Could they have been more pronounced
 during the giant eruptions of the nineteenth century? There is no reason to believe
  so, as $\eta$ Car was much brighter in the optical than it is now, diminishing the 
  contrast between the quasi-periodic maxima and the 
underlying stellar light. However, this is unsafe terrain, as we do not know 
what mechanism produced the pronounced peaks during the giant eruption.

Recent estimates of the Homunculus mass suggest that more than
$10\,M_{\odot}$ of material was ejected during the great eruption
\citep{SGH03_IR}. Given the larger mass-loss that occurred in 
the giant eruption it is likely that the orbital parameters changed substantially
during the great eruption, and thus it is not surprising there is not
a one to one correspondence between orbital phase and the pronounced
peaks observed in 1827--1843, even if they were a result of a binary interaction near 
periastron.

\section{ DISCUSSION AND CONCLUSIONS}
\label{discus}

We presented a homogeneous set of spectra covering the
events \#9 (1992.42), \#10 (1997.95) and \#11 (2003.49). We
derived the period by measuring intensities of narrow lines
(Weigelt blobs) and broad emission lines (stellar wind) and radial velocity
variations from broad line components. These data,
and others collected from the literature, enable an accurate
determination of the period P$_{\rm{pres}}$ = 2022.7$\pm$1.3 d. An average
period encompassing the past eleven cycles was found
to be P$_{\rm{avg}}$ = 2020$\pm$ 4 d, compatible with the present day
period. The period change is smaller than
1.5 d cycle$^{-1}$ along the last half century. 

It is difficult to imagine any mechanism other than
orbital motion which could maintain such a high stability,
while allowing individual features to show the distinct light curves
that are observed. No luminous unstable star is expected to 
follow such a precise clock. Even if it is conceivable that a shell ejection
 could be involved in the periodic events, it should be triggered by the 
periastron passage, when the secondary star approaches the 
primary to a few stellar radii. However, in an incoming paper we 
show that the event starts when the secondary star is still 
near apastron.

We must expect that the period is drifting,
since the stars are losing mass, they interact strongly as they get
very close at periastron and the primary star appears to be
a fast rotator. The period changes have been $\Delta$P/P $<$ 1/1000  
along the last 60 years, which is consistent with the observed mass loss rate.

The situation is different for the eruptions of 1843 and 1890, when a considerable 
amount of matter was removed from the primary star in brief episodes. However, 
the relation between the pronounced peaks observed
during the great eruption and periastron passages continues
to be unknown. This is because the coincidence is
not perfect, and because the peaks might not be strictly
periodic but could still be associated with periastron passages,
as seen presently in the near-infrared light curves.

From the disappearance of the He\,{\sc i}~$\lambda6678$ narrow component
we determined the epoch of the start of the minimum to beT$_{0}$ = JD~2,452,819.8. 
The procedure to define
the minimum requires fitting and extrapolating the line intensity
variation along the descending part of the line intensity curve, in the two weeks
 preceding the minimum intensity. Because of this definition,
and since the starting time of the minimum is 
different from line to line, this definition is
arbitrary and has no physical meaning. However, it is robust
and demands only a few observations along the $\sim$3 weeks
before the complete disappearance of the feature. Importantly,
the time delay for all other features to reach the minimum is well known.

The next minimum is predicted to start on 2009 January 11 ($\pm$2 d).  
 This will the best event since 1948 for
ground-based observations, since its central core fits entirely in the good
observing season. The next favorable event will not occur before
2020. In order to improve the results presented in this work, daily observations 
should be made along a month starting on 2008 December 20.

\section{Acknowledgments}
 We thank J.E. Steiner and T. Gull for their comments on the draft. A.D., J.H.G. 
and M.T. thank to FAPESP and CNPq for continuing support. Financial 
support from PIP-CONICET No. 5697 is acknowledged by J.A.. DJH acknowledges
partial support from HST AR-10957.

\bsp

\label{lastpage}

\end{document}